\documentclass[prd,letterpaper,twocolumn,tightenlines,superscriptaddress,
showpacs,psfig,showpacs,showkeys]{revtex4}
\usepackage{amsmath,amssymb,amsfonts,graphicx,dcolumn}
\usepackage{epstopdf}
\usepackage{color}
\usepackage[utf8]{inputenc}
\usepackage{amsthm}
\usepackage{fontenc}
\usepackage[dvips]{hyperref}
\usepackage{bm}
\RequirePackage{slashed}
\usepackage{cancel}
\usepackage[normalem]{ulem}
\newcommand{\ie}{\textsl{i.e. }}

\usepackage{array,booktabs}
\newcolumntype{C}{>{$\displaystyle}c<{$}}

\begin{document}

\pacs{}
\keywords{}

\twocolumngrid
\title{
Parton correlations in
same-sign $W$ pair production 
via double parton scattering
at the LHC
}

\author{Federico Alberto Ceccopieri}
\affiliation{ Dipartimento di Fisica e Geologia,
Universit\`a degli Studi di Perugia and Istituto Nazionale di Fisica Nucleare,
Sezione di Perugia, via A. Pascoli, I - 06123 Perugia, Italy}
\affiliation{IFPA, Université de Li\`ege, B4000, Li\`ege, Belgium}
\email{federico.alberto.ceccopieri@cern.ch}	
\author{Matteo Rinaldi}
\affiliation{Departament de Fisica Te\`orica, Universitat de Val\`encia
and Institut de Fisica Corpuscular, Consejo Superior de Investigaciones
Cient\'{\i}ficas, 46100 Burjassot (Val\`encia), Spain}
\author{Sergio Scopetta}
\affiliation{ Dipartimento di Fisica e Geologia,
Universit\`a degli Studi di Perugia and Istituto Nazionale di Fisica Nucleare,
Sezione di Perugia, via A. Pascoli, I - 06123 Perugia, Italy}
\date{\today}

\begin{abstract}
\vspace{0.5cm}
Same-sign $W$ boson pairs production is a promising channel to look for
signatures of double parton interactions at the LHC.
The corresponding cross section 
has been calculated
by using 
double parton distribution functions, encoding
two parton correlations, evaluated in
a Light-Front quark model.
The obtained result is in line with previous estimates
which make use of an external parameter, the so called
effective cross section, not necessary in our approach. 
The possibility to observe for the first time
two-parton correlations, 
in the next LHC runs, has been established.

\end{abstract}

\maketitle
 
It is known since a long time that a 
proper description of final states in hadronic collisions 
requires the inclusion of
processes where more than one pair of partons 
participate in a single hadronic collision,
the so-called multiple partonic interactions 
(MPI)~\cite{paver,sjostrand}.
Due to LHC operation, the wide subject of 
MPI is having in these years a renewed interest~\cite{diehl_1}.
At low transverse momenta, MPI enhance particle production and 
affect particle multiplicities and energy flows.
The effect of MPI is present also in hard scattering processes.
In this letter, we are interested in double parton scattering (DPS),
in which parton pairs from two hadrons interact between each other,
and both collisions are hard enough to apply perturbative techniques.
While these processes need to be well controlled since
they could represent a background to New Physics searches, the main 
focus of this work is the sensitivity of DPS to relevant features
of the non-perturbative nucleon structure, not accessible otherwise.
In particular, DPS cross section  
depends on non-perturbative quantities, the so-called double parton 
distribution functions (dPDFs). The latter
represent the number density of parton pairs with longitudinal 
fractional momenta $x_1,x_2$, at a relative transverse distance
${\vec{b}_\perp}$. If extracted from data, dPDFs would offer for the first time 
the opportunity to investigate two-parton correlations,
as noticed long time ago \cite{calucci}.
Since dPDFs are two-body distributions, this 
information is different and
complementary to the one encoded in 
one-body distributions, such as ordinary
and generalized parton distributions 
\cite{gpdtmd}.
The present letter aims at establishing to what extent
this novel information can be accessed in the next runs of LHC, looking at
a specific final state, namely, the production of a pair
of $W$ bosons with the same-sign ($ssWW$).
In fact, this channel has been found to be promising for
DPS observation~\cite{Kulesza:1999zh,Maina:2009sj,Gaunt:2010pi},
since single parton scattering (SPS) at tree-level starts contributing
to higher order in the strong coupling~\cite{Campbell:2011bn}. For such reasons,
diboson production via DPS has been
theoretically investigated in 
detail~\cite{Goebel,Mekhfi:1983az,DG_Glauber,Kasemets:2012pr}. 

Let us define now the quantities we are going to calculate.
If final states $A$ and $B$ are produced in a DPS process,
the corresponding cross section 
can be sketched as \cite{paver} 
\begin{multline}
d \sigma^{AB}_{DPS} = {m \over 2} 
\sum_{i,j,k,l} \int d \vec{b}_\perp 
D_{ij}(x_1,x_2;\vec{b}_\perp)  \\
\times D_{kl}(x_3,x_4;\vec{b}_\perp) \, d \hat \sigma_{ik}^A \, d \hat 
\sigma_{jl}^B\,,
\label{uno}
\end{multline}
where $m=1$
if $A$ and $B$ are identical and $m=2$ otherwise,
$i,j,k,l = \{q, \bar q, g \}$ are the parton species contributing
to the final states $A (B)$.
In Eq. (\ref{uno}) and in the following,
$d \sigma$ is used for the cross section, differential in the relevant
variables.  
The functions $D_{ij}$ in Eq.~(\ref{uno}) 
are the dPDFs
which depend additionally 
on factorization scales $\mu_{A(B)}$,
$D_{ij}(x_1,x_2; \vec{b}_\perp,\mu_A,\mu_B)$.
\indent
To date, dPDFs are very poorly known, so that  it has 
been useful
to describe DPS cross section independently of the dPDFs concept, using the
approximation:
\begin{eqnarray}
d\sigma^{AB}_{DPS}  \simeq \dfrac{m}{2} d\sigma_{SPS}^A \dfrac{ 
d\sigma_{SPS}^B}{\sigma_{eff}}\,,
\label{sigma_eff_exp}
\end{eqnarray}
where $d\sigma^{A}_{SPS}$ is the SPS cross section with final 
state $A$:
\begin{eqnarray}
\label{s_single}
d\sigma_{SPS}^{A}= 
\sum_{i,k}  f_{i} (x_{1},\mu_A)
f_{k}(x_3,\mu_A)\,
d\hat \sigma_{ik}^{A}(x_1,x_3,\mu_A)\,.
\end{eqnarray}
In Eq.~(\ref{s_single}) $f_{i(j)}$ are parton distribution functions (PDFs) and an analogous expression holds 
for the final state $B$.
The physical meaning of Eq.~(\ref{sigma_eff_exp}) is that, 
once the process $A$ has occured with cross section $\sigma_{SPS}^A$, 
the ratio $\sigma_{SPS}^B / \sigma_{eff}$ represents the probability of process B 
to occur.
A constant value of $\sigma_{eff}$
has been assumed in 
all the experimental analyses performed so far,
so that the technical implementation of Eq.~(\ref{sigma_eff_exp}) is rather easy.
In this way, different collaborations have extracted values of 
$\sigma_{eff}$ which are consistent
within errors,
irrespective of center-of-mass energy of the hadronic 
collisions and of the final state considered. 
A comprehensive compilation 
of experimental results on $\sigma_{eff}$ is reported in Ref. \cite{ATLAS_2016_4jet},
where the latest DPS measurement in the four jet final state is presented.

To understand the approximation leading to Eq. (\ref{sigma_eff_exp}) from Eq. (\ref{uno}), let us
write dPDFs in the latter in a fully 
factorized form:
\begin{eqnarray}
\label{fact}
D_{ij}(x_1,x_2,\mu_A,\mu_B,\vec{b}_\perp)  = 
f_{i} (x_{1},\mu_{A})
f_{j}(x_{2},\mu_{B})\,
T(\vec{b}_\perp)~,
\end{eqnarray}
where the function $T(\vec{b}_\perp)$, 
describes the probability to have two partons 
at a transverse distance $\vec b_\perp$.
Then, inserting Eq. (\ref{fact})
into Eq. (\ref{uno}), one obtains Eq. 
$\sigma_{eff}$, 
(\ref{sigma_eff_exp}), as follows
\begin{equation}
\sigma_{eff}^{-1} =  \int d \vec{b}_\perp [
T(\vec{b}_\perp)
]^2~,
\label{bprofile}
\end{equation}
with $T(\vec{b}_\perp)$ controlling the
double parton interaction rate.
It is clear that, as a consequence of the approximation~(\ref{fact}),
$\sigma_{eff}$ does not show any dependence on parton fractional momenta, 
hard scales or parton species.

Actually, if factorized expressions are not used,
$\sigma_{eff}$
depends on longitudinal momenta.
Since dPDFs are basically unknown, and only
sum rules relating them to PDFs are available
~\cite{GS09,ceccopieri2},  model calculations, developed at low energy, but able 
to reproduce relevant features of nucleon parton 
structure,  can be useful and have been proposed.
In such model calculations, factorized structures, Eq. (\ref{fact}), do not arise,
and $\sigma_{eff}$ depends
non-trivially on longitudinal momenta.
In particular, this was found in a Light-Front (LF) Poincar\'e covariant constituent quark model (CQM),
reproducing the sum rules of 
dPDFs ~\cite{noi2,nois}, as well as
in a holographic approach~\cite{Traini:2016jru}.
In this letter
we will 
evaluate
DPS cross sections,
using different models of dPDFs,
to establish wether forthcoming LHC data
will exhibit (for the considered final state)
such features, not yet seen 
in the present uncertain experimental scenario. 

Let us now  summarise our  calculation.
We first consider the SPS $W^{\pm}$ production and 
subsequent decay into muon at center-of-mass energy $\sqrt{s}$:
\begin{equation}
pp\rightarrow W^\pm(\rightarrow \mu^\pm \; \nu_\mu^{\hspace{-0.3cm}(-)})X\,,
\end{equation}
indicating with $\sigma^\pm$ the corresponding cross sections. 
Defining quarks according to their charge, \ie $D=d,s,b$ and $U=u,c,t$, 
we consider the following partonic subprocesses 
\begin{eqnarray}
U(p_a) \bar{D}(p_b) &\rightarrow& \mu^+(p_\mu) \nu_\mu(p_{\nu})\,,	\\ 
D(p_a) \bar{U}(p_b) &\rightarrow& \mu^-(p_\mu) \bar{\nu}_\mu(p_{\nu})\,, 
\end{eqnarray}
where particle four-momenta are indicated in parenthesis.
Differential cross sections are calculated in terms of the muon transverse 
momentum 
$p_T=|\vec{p}_T|$ and pseudorapidity $\eta_\mu$,  defined in the hadronic center-of-mass frame. 
The partonic Lorentz invariants $\hat{u}$ and $\hat{t}$, in terms of these variables, read 
\begin{eqnarray}
\hat{t}=(p_a-p_\mu)^2&=&-x_a \sqrt{s} p_T e^{-\eta_\mu}\,, \nonumber\\
\hat{u}=(p_b-p_\mu)^2&=&-x_b \sqrt{s} p_T e^{\eta_\mu}\,,
\end{eqnarray}
from which parton fractional momenta can be calculated as 
\begin{equation}
x_a=e^{\eta_\mu} \frac{M_W}{\sqrt{s}}(A \pm B)\,, \;\;
x_b=e^{-\eta_\mu} \frac{M_W}{\sqrt{s}}(A \mp B)\,,
\label{xaxb}
\end{equation}
with $A=M_W/(2p_T)$, $B=\sqrt{A^2-1}$ and $M_W$ the $W$-boson mass. The 
unobserved neutrino causes an under-determination 
of the $W$-rapidity and, in turns, the twofold ambiguity in Eq.~(\ref{xaxb}).  
Cross sections are evaluated in the narrow width approximation, \ie at fixed $\hat{s}=(p_a+p_b)^2=M_W^2$, and
read
\begin{eqnarray}
\label{Wprod_CS}
&& \frac{d^2 \sigma^{pp\rightarrow W^+ (\rightarrow \mu^+ \nu)X}}{d\eta dp_T}=\frac{G_F^2}{6 s \Gamma_W} 
\frac{V_{U\bar{D}}^2}{\sqrt{A^2-1}}  \\
&& \times \Big[ f_U(x_a, \mu_F) f_{\bar{D}}(x_b,\mu_F) \hat{t}^2 + 
f_{\bar{D}}(x_a,\mu_F)f_U(x_b,\mu_F) \hat{u}^2 \Big]\,, \nonumber
\end{eqnarray}
where $G_F$ is the Fermi constant, $\Gamma_W$ the $W$ boson decay width and $V_{ij}$ the CKM matrix elements
whose values are taken from Ref.~\cite{PDG}. 
The $\sigma^-$ cross section is obtained exchanging 
$U \leftrightarrow D$ and $\hat{t} \leftrightarrow \hat{u}$ in Eq.~(\ref{Wprod_CS}).
The PDFs appearing in Eq.~(\ref{Wprod_CS}) are evaluated at a factorization scale $\mu_F=M_W$
and therefore PDFs from CQM calculations, related to low momentum scales,  need to be properly evolved.
The evolution is performed at LO by using DGLAP equations. 
We adopt a variable flavour number scheme 
and parameters as in LO version of \texttt{MSTW08} distribution~\cite{MSTW08}. 
In particular  
heavy quark masses are set to $m_c=1.4$ GeV and $m_b=4.75$ GeV and the one-loop running coupling 
is fixed at $Z$-boson mass scale to be $\alpha_s^{(n_f=5)}(M_Z^2) = 0.13939$~\cite{MSTW08}.
\begin{figure}[t]
\includegraphics[scale=0.90]{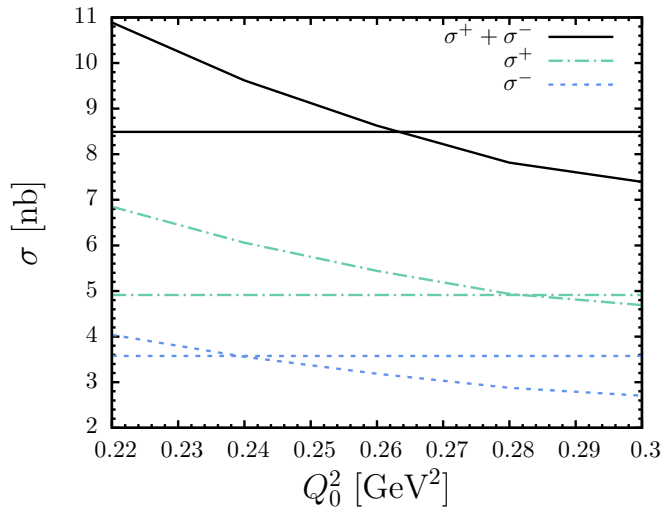}
\caption{\textsl{W production cross sections as predicted by LF PDFs as a function of $Q_0^2$, compared to
\texttt{DYNNLO} predictions (straight lines) at LO by using LO \texttt{MSTW08} parton distributions in the fiducial 
region indicated in the text. }}
\label{scale_fixing}
\end{figure}
For PDFs provided by the LF CQM one has  
\begin{equation}
\label{ic_single}
f_{d}(x,Q_0^2)=1/2 f_{u}(x,Q_0^2)\,,
\end{equation}
at the hadronic scale $Q_0^2$, where three valence quarks carry all proton momentum.
Since this scale is generally located in the infrared regime, PDFs evolution and corresponding cross sections 
are very sensitive to its choice. 
defined.
Therefore, in the present paper, $Q_0^2$  is fixed requiring 
that 
$\sigma^+$ and $\sigma^-$, calculated by using evolved LF PDFs, match
the corresponding predictions obtained with the \texttt{DYNNLO} code~\cite{DYNNLO} at LO by using \texttt{MSTW08} PDFs~\cite{MSTW08}.
For both simulations we set $\sqrt{s}=13$ TeV and define the muon fiducial 
phase space in SPS 
to be $p_T^\mu>20$ GeV and $|\eta^{\mu}|<2.4$.
As shown in Fig.~\ref{scale_fixing}, considering the cross section summed over the $W$ boson charge, 
this procedure locates the central value of the initial scale at $Q_0^2=0.26$ Ge$\mbox{V}^2$
(where $\alpha_s(Q_0^2)=1.99$). We note that, for a given value of $Q_0^2$, a simultaneous description of 
$\sigma^+$ and $\sigma^-$ can not be achieved, a fact which is ascribed to the model assumption for  
PDFs in Eq.~(\ref{ic_single}) and it is an example of typical drawback of PDFs CQM calculations. 
In order to take into account this deficiency, we assign a theoretical 
error to $Q_0^2$, allowing it to vary in the range $0.24 < 
Q_0^2 < 0.28 
\;\mbox{GeV}^2$, 
where the limits are fixed requiring that cross sections
obtained via LF model reproduce $\sigma^+$ and $\sigma^-$ predicted by \texttt{DYNNLO} 
(straight lines in Fig.~\ref{scale_fixing}).
Having fixed $Q_0^2$ in SPS processes and being dPDFs obtained within the same 
LF model adopted for PDFs,  we can 
use the same $Q_0^2$ range for dPDFs. In this way the estimate of DPS cross sections
does not require additional parameters. Double PDFs in the LF model are defined at $Q_0^2$ as
\cite{noi2}
\begin{equation}
\label{ic_dpdfs}
f_{du}=f_{ud}=f_{uu}(x_1,x_2,Q_0^2,\vec{b}_\perp)\,. 
\end{equation}
\vskip -.1cm
At this scale, when integrated over $\vec{b}_\perp$, dPDFs satisfy number and 
momentum sum rules~\cite{GS09}.
Their perturbative QCD evolution is presently known only at leading
logarithmic accuracy~\cite{snigirev03,ceccopieri1}, however the presence of 
the so-called  inhomogeneous term in the evolution equations is still under investigation~\cite{diehl_1,ceccopieri2,blok_1}.
In the present paper dPDFs are evolved with the same scheme and parameters used for PDFs but using 
homogeneous evolution equations valid at fixed values of $\vec{b}_\perp$~\cite{diehl_1,dkk}.
The DPS cross section, Eq.~(\ref{uno}), in the $ssWW$ channel reads
\begin{multline}
\frac{d^4 \sigma^{pp\rightarrow \mu^\pm\mu^\pm X}}{d\eta_1 dp_{T,1} d\eta_2 dp_{T,2} }= \displaystyle  \sum_{i,k,j,l}
\frac{1}{2} \int d^2 \vec{b}_\perp \\ \times D_{ij}(x_1,x_2,\vec{b}_\perp, M_W) 
D_{kl}(x_3,x_4,\vec{b}_\perp, M_W) \\
\times \frac{d^2 \sigma_{ik}^{pp\rightarrow \mu^\pm X}}{d\eta_1 dp_{T,1}}
\frac{d^2 \sigma_{jl}^{pp\rightarrow \mu^\pm X}}{d\eta_2 dp_{T,2}} \mathcal{I}(\eta_i,p_{T,i})\,.
\label{DPS_cs}
\end{multline}
The function $\mathcal{I}(\eta_i,p_{T,i})$ in Eq.~(\ref{DPS_cs})
implements the kinematical cuts reported in Tab.~(\ref{FXS})
which we mutuate from the 8 TeV analysis of Ref.~\cite{CMS_AN}.
In Eq.~(\ref{DPS_cs}) we are neglecting the supposed small contributions  
coming from longitudinally polarized dPDFs~\cite{Kasemets:2012pr}.
Eq.~(\ref{DPS_cs}) will be evaluated with three different
models of dPDFs described in the following in order of increasing complexity. 
In the simplest one, called MSTW, dPDFs are parameterized as products
of {\texttt{MSWT08}} PDFs according to Eq.~(\ref{fact}).
{\renewcommand\arraystretch{1.5} 
\begin{table}[t]
\begin{center}
\begin{tabular}{c}   \hline  \hline
$pp$, $\sqrt{s}=13$ TeV\\ 
$p_{T,\mu}^{leading}>20$ GeV, \;\; $p_{T,\mu}^{subleading}>10$ GeV\\
$|p_{T,\mu}^{leading}|+|p_{T,\mu}^{subleading}|>45$ GeV\\
$|\eta_{\mu}|<2.4$ \\
20 GeV $<$ $M_{inv}$ $<$ 75 GeV or $M_{inv}$ $>$ 105 GeV \\ \hline 
\end{tabular}
\caption{\textsl{Fiducial DPS phase space used in the analysis. }}
\label{FXS}
\end{center}
\end{table}}
In the second one, the so-called GS09 model~\cite{GS09},
the factorized form Eq.~(\ref{fact}), properly corrected to fulfill dPDFs sum rules,
is assumed only at a momentum scale $Q_0^2$.
Such initial conditions are evolved with dPDFs evolution
equations with the inhomogenous term included~\cite{snigirev03,ceccopieri1}.
Therefore, with respect to model MSTW, GS09 takes into account additional
perturbative correlations~\cite{calucci,GS09,dkk,noinew}. The
DPS cross section based on MSTW and GS09 models can be evaluated 
only assuming a constant $\sigma_{eff}$ in 
Eq.~(\ref{sigma_eff_exp}).
In the present work, we will use, as a reference value, 
$\bar \sigma_{eff}=17.8 \pm 4.2$ mb,
which is the average of two recent
extractions~\cite{S5,S6} in the $W$-boson plus dijet final state, the latter
being the closest to the one considered here.
The only available information in this channel
is a lower limit, $\sigma_{eff}>5.91$ mb at 95\% confidence level, recently obtained
with an integrated luminosity ${\cal L}$=19.7 fb$^{-1}$ at $\sqrt{s}=$ 8 TeV \cite{CMS_AN}.

In the last model~\cite{noi2}, called QM,
dPDFs have been evaluated within the LF framework,
generalizing the approach of Ref.~\cite{LF2} for the calculation of 
PDFs. 
As a result, fully correlated dPDFs are obtained~\cite{noi2}. In such a model,
longitudinal and transverse correlations are generated among 
valence quarks 
and propagated by perturbative evolution to sea quarks and gluons dPDFs.
The use of this model in the present analysis is particularly relevant.
First of all, within this model, 
the DPS cross section
can be calculated using Eq.~(\ref{uno}), without any assumption
on $\sigma_{eff}$, at variance with MSTW and GS09. Moreover, the 
simultaneous use of 
single and double PDFs obtained from the same LF dynamics, 
allows one to investigate the role of 
parton correlations
on potentially sensitive observables.
\setlength{\extrarowheight}{0.2cm}
\begin{table}[t]
\begin{center}
\begin{tabular}{|c|c|c|c|}  \hline  \hline 
dPDFs & $\sigma^{++}+ \sigma^{--}$ [fb] \\ [0.2cm] \hline
MSTW&   $0.77^{\hspace{0.1cm}+0.23}_{\hspace{0.1cm}-0.21}$ ($\delta \mu_F$) $\phantom{}^{+0.18}_{-0.18}$ ($ \delta \bar{\sigma}_{eff}$) \\ [0.2cm]  \hline
GS09&   $0.82^{\hspace{0.1cm}+0.24}_{\hspace{0.1cm}-0.26}$ ($\delta \mu_F$) $\phantom{}^{+0.19}_{-0.19}$ ($ \delta \bar{\sigma}_{eff}$) \\ [0.2cm]  \hline
QM  &   $0.69^{\hspace{0.1cm}+0.18}_{\hspace{0.1cm}-0.18}$ ($\delta \mu_F$) $\phantom{}^{+0.12}_{-0.16}$ ($\delta Q_0$)  \\ [0.2cm] \hline 
\end{tabular}
\caption{\textsl{Model predictions for
$W$-charge summed cross sections in fiducial region in Tab.~(\ref{FXS}).}}
\label{ssWW_ref}
\end{center}
\end{table}
\setlength{\extrarowheight}{0.0cm}
\begin{table}[t]
\begin{center}
\begin{tabular}{|c|c|c|c|}   \hline \hline
dPDFs & $\sigma^{++}$ [fb] & $\sigma^{--}$ [fb] & $\sigma^{++}/\sigma^{--}$ \\ \hline
GS09      & 0.54 &  0.28  & 1.9    \\
QM        & 0.53 &  0.16  & 3.4    \\ \hline
GS09/QM   & 1.01 &  1.78  &  -      \\ \hline 
\end{tabular}
\caption{\textsl{Ratio of cross sections for same sign muons production in fiducial region.}}
\label{ration_pdf_dpd}
\end{center}
\end{table}
\begin{figure}[t]
\includegraphics[scale=0.90]{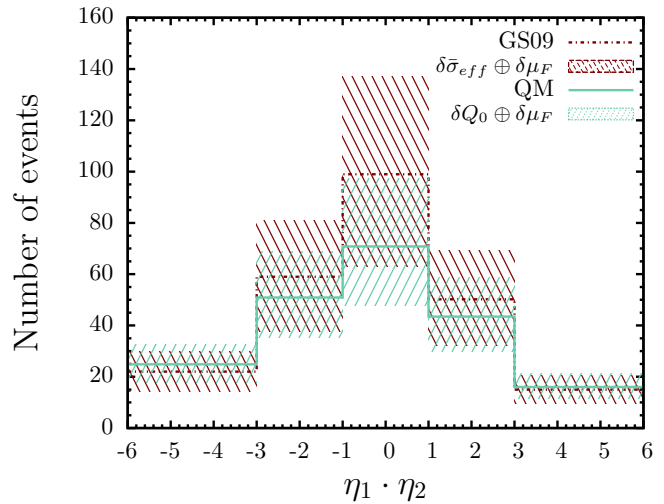}
\caption{\textsl{Number of expected events with $\mathcal{L}^{-1}=300$ f$\mbox{b}^{-1}$
as a function of product of muons rapidities.}}
\label{dsdpeta_nev}
\end{figure}
Theoretical systematic errors are associated to our predictions as follows.
Uncertainties related to missing higher order corrections,  denoted by $\delta \mu_F$, are estimated for all models,
varying $\mu_F$ in the range $0.5 M_W <\mu_F<2.0 M_W$;
the ones due  $Q_0$-fixing, denoted by $\delta Q_0$,  are given by varying this parameter in the range $0.24 < Q_0^2 < 0.28 \;\mbox{GeV}^2$.
A further error, $\delta \bar{\sigma}_{eff}$, is assigned to MSTW and GS09 predictions, due to $\bar{\sigma}_{eff}$ 
uncertainty.
In Tab.~\ref{ssWW_ref} we report DPS cross sections, integrated in the fiducial volume, 
evaluated using the above models.
Predictions based on MSTW and GS09 are close, while QM one is smaller by around 15\%,
although they are all consistent within errors.

For all the considered models, cross sections rise as $\mu_F$ increases, 
an effect induced by the sea quark growth at $\langle x \rangle \sim 10^{-2}$ (typical of this process).
We have estimated that, if the integrated luminosity ${\cal L}$ is greater than 300 fb$^{-1}$,
the central values of the QM and GS09 predictions can be discriminated.
It is worth noting that, if the measurement were performed also in the $e\mu$ 
($e\mu$+$ee$) channel, 
the number of signal events would increase by a factor three 
(four).

In Tab.~\ref{ration_pdf_dpd}, predictions of models GS09 and QM
for default values of parameters of charged $ssWW$ cross sections (indicated by $\sigma^{--}$ and $\sigma^{++}$) 
integrated in the fiducial volume, are compared. While agreement between model predictions is found for $\sigma^{++}$, 
a rather smaller $\sigma^{--}$ is obtained in model QM,  due to the assumption in Eq.(10).
The ratio  $\sigma^{--}$ and $\sigma^{++}$ is therefore a suitable observable to 
investigate the flavor structure of dPDFs.

In order to analyze correlations encoded in dPDFs, we consider the differential cross section in the variable  
$\eta_1 \cdot \eta_2$ which, neglecting the boost generated by $W$-decay into leptons, can be approximated via
Eqs.~(\ref{xaxb}) as
\begin{equation}
\eta_1 \cdot \eta_2  \simeq \frac{1}{4} \ln \frac{x_1}{x_3} \ln \frac{x_2}{x_4}\,,
\label{p}
\end{equation}
where fractional momenta are subject to the invariant mass constraint $x_1 x_3= x_2 x_4= M_W^2/s$.
The result, converted into per-bin number of events
assuming an integrated luminosity $\mathcal{L}=300$ f$\mbox{b}^{-1}$, is presented in Fig.~\ref{dsdpeta_nev}.
The maximun is located at $\eta_1 \cdot \eta_2 \sim 0$, where annihilating partons 
equally share the momentum fractions, $x\sim M_W/\sqrt{s}$, in at least one scattering.
At large and positive (negative) values of $\eta_1 \cdot \eta_2$, muons are produced
in the same (opposite) emisphere and the fast drop of the cross section is associated to the 
fall off of dPDFs as one ($\eta_1 \cdot \eta_2 \ll0$) or both ($\eta_1 \cdot \eta_2 \gg0$) partons in the same proton approach the large $x$ limit. 
We note that predictions based on GS09 and QM models show a rather similar shape and are compatible 
within their sizeable errors.
To deal with such large uncertainties, differential cross sections, normalized to the total ones
(Tab.~\ref{ssWW_ref}), may be considered. 
In this way, the predictions of MSTW and GS09 
models do not depend any more on the choice of
$\sigma_{eff}$ and the related error cancels. 
Moreover, for model QM, we verified that the scale variations
$\delta{\mu_F}$ and $\delta{Q_0}$, acting basically on normalizations, 
almost cancel in the ratio.
A shape comparison can then be used to discriminate 
among models and their factorized
structure. In the present analysis, 
however, we prefer to discuss the effects of correlations on a more familiar 
quantity, $\sigma_{eff}$.

\begin{figure}[t]
\includegraphics[scale=0.90]{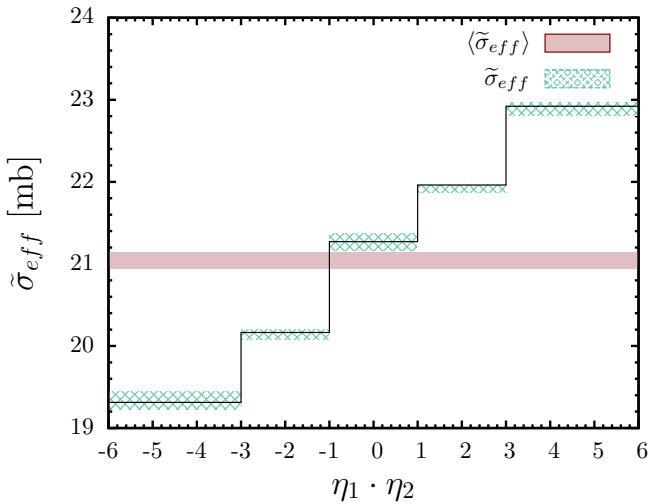}
\caption{\textsl{$\widetilde{\sigma}_{eff}$ and $\langle 
\widetilde{\sigma}_{eff} \rangle$ as a 
function of product of muon rapidities. The error band represents scale 
variations added in quadrature.}}
\label{seff}
\end{figure}

To this aim we use the LF approach for both PDFs and dPDFs to evaluate SPS and DPS differential cross sections, 
Eqs. (11) and (14), respectively, integrated in bins of $\eta_1 
\cdot \eta_2$. With these ingredients we obtain, through Eq. (2), 
a prediction for $\sigma_{eff}$ intrinsic to the LF model,  
called hereafter $\widetilde \sigma_{eff}$.
If a corresponding procedure is performed on cross sections integrated in 
the
fiducial volume, one obtains the constant value
\begin{equation}
\langle \widetilde{\sigma}_{eff} \rangle = 21.04^{\hspace{0.1cm}+0.07}_{\hspace{0.1cm}-0.07} \, (\delta Q_0) 
\, \phantom{}^{+0.06}_{-0.07} (\delta \mu_F) \; \mbox{mb}~.
\label{s_hat}
\end{equation}
This value is compatible, within errors, 
with $\bar{\sigma}_{eff}$  experimentally determined.
Both $\widetilde{\sigma}_{eff}$ and $\langle \widetilde{\sigma}_{eff} \rangle$ 
are shown in Fig.~\ref{seff}
and, being ratios, are both stable against $\mu_F$ and $Q_0$ variations.
The departure of $\widetilde{\sigma}_{eff}$ from a constant value is a measure
of two parton correlations in the proton. 
These are primarily correlations in longitudinal momenta  
but, as shown using the fully correlated model QM, 
they are related to
the ones in transverse space in an irreducible way~\cite{Rinaldi_Ceccopieri}. 
We have estimated that this departure could be appreciated 
with an integrated luminosity ${\cal L}$ 
of around 1000 fb$^{-1}$, at 68 \% confidence level, 
reachable in the planned LHC runs. 
Our 
conclusion is that the extraction of 
this observable in
bins of $\eta_1 \cdot \eta_2$
is a convenient strategy to look for parton correlations.
\indent Summarising, we have calculated $ssWW$ cross sections in a LF model for dPDFs,
carefully estimating the corresponding uncertainties. 
Our predictions, completely intrinsic to the approach,  are in line with those obtained by other approaches 
which make use of the external parameter $\sigma_{eff}$. This indicates that the model is able to catch 
the transverse structure of the DPS process. 
Furthermore, we have established that, in this specific final state, transverse and longitudinal correlations, embodied 
in dPDFs, could be observed in the next LHC runs.
\\ \indent This work is supported in part through the project 
``Hadron Physics at the LHC: looking for signatures of
multiple parton interactions and quark gluon plasma formation (Gossip project)",
funded by the ``Fondo ricerca di base di Ateneo" of the Perugia University. 
This work is also supported in part by Mineco under contract 
FPA2013-47443-C2-1-P and SEV-2014-0398.
We  warmly thank Livio Fan\`o, Marco Traini and
Vicente Vento for 
many useful discussions.

\end{document}